\documentclass[letter,12pt]{extarticle}

\usepackage{multicol}
\usepackage{amsmath}
\usepackage[T1]{fontenc}
\usepackage[dvipsnames]{xcolor}
\usepackage[section]{placeins}
\usepackage{xcolor}
\usepackage[toc,page,header]{appendix}

\usepackage{graphicx}
\usepackage[font=small,labelfont=bf]{caption}
\usepackage{multirow}
\usepackage{caption}
\usepackage{enumitem}
\usepackage{subcaption}
\usepackage[utf8]{inputenc}
\usepackage{chngcntr}
\usepackage[style = numeric-comp, sorting = none, maxnames=10 ]{biblatex}
\usepackage{authblk}

\bibliography{biblio.bib}
\usepackage[width=210.00mm, height=297.00mm, left=2cm, right=2cm, top=2cm, bottom=2.5cm]{geometry}
\usepackage{ amssymb }
\usepackage{verbatim}
\usepackage{caption}
\usepackage{subcaption}

\title{CREIME -- A \textit{C}onvolutional \textit{R}ecurrent model for \textit{E}arthquake \textit{I}dentification and \textit{M}agnitude \textit{E}stimation }

\date{}

\author[1,4]{Megha Chakraborty}
\author[1,6]{Darius Fenner}
\author[1]{Wei Li}
\author[1,2]{Johannes Faber}
\author[1,2,3]{Kai Zhou}
\author[1,4]{Georg Rümpker}
\author[1,2,3,5]{Horst Stoecker}
\author[1,4 *]{Nishtha Srivastava}

\affil[1]{Frankfurt Institute for Advanced Studies, 60438 Frankfurt am Main, Germany}
\affil[2]{Institute for Theoretical Physics, Goethe Universität, 60438 Frankfurt am Main, Germany}
\affil[3]{Xidian-FIAS international Joint Research Center, Giersch Science Center,
60438 Frankfurt am Main, Germany}
\affil[4]{Institute of Geosciences, Goethe-University Frankfurt, 60438 Frankfurt am Main, Germany}
\affil[5]{GSI Helmholtzzentrum für Schwerionenforschung GmbH, 64291 Darmstadt, Germany}
\affil[6]{Johannes Gutenberg-Universität Mainz, 55122, Mainz,Germany}
\affil[*]{srivastava@fias.uni-frankfurt.de}

\renewcommand{\baselinestretch}{1.5} 

\begin{document}
\renewcommand{\baselinestretch}{1.2} 
\maketitle

\renewcommand{\baselinestretch}{1.5} 
\begin{abstract}
The detection and rapid characterisation of earthquake parameters such as magnitude are of prime importance in seismology, particularly in applications such as Earthquake Early Warning (EEW). Traditionally, algorithms such as short-term average/long-term average (STA/LTA) are used for event detection, while frequency or amplitude domain parameters calculated from 1-3 seconds of first P-arrival data are sometimes used to provide a first estimate of (body-wave) magnitude. Owing to extensive involvement of human experts in parameter determination, these approaches are often found to be insufficient. Moreover, these methods are sensitive to the signal-to-noise ratio and may often lead to false or missed alarms depending on the choice of parameters. We, therefore, propose a multi-tasking deep learning model -- the \textbf{C}onvolutional \textbf{R}ecurrent model for \textbf{E}arthquake \textbf{I}dentification and \textbf{M}agnitude \textbf{E}stimation (CREIME) that: (i) detects the first earthquake signal, from background seismic noise, (ii) determines first P-arrival time as well as (iii) estimates the magnitude using the raw 3-component waveform data from a single station as model input. Considering, that speed is of the essence in EEW, we use up to two seconds of P-wave information which, to the best of our knowledge, is a significantly smaller data window (5 second window with up to 2 seconds of P-wave data) compared to the previous studies. To examine the robustness of CREIME we test it on two independent datasets and find that it achieves an average accuracy of 98\% for event-vs-noise discrimination and is able to estimate first P-arrival time and local magnitude with average root mean squared errors of 0.13 seconds and 0.65 units, respectively. We also compare CREIME architecture with the architectures of other baseline models, by training them on the same data, and also with traditional algorithms such as STA/LTA, and show that our architecture outperforms these methods.
    
\end{abstract}

\section{Introduction}
According to its original definition \cite{richter} the magnitude of an earthquake is the logarithm of the maximum trace amplitude expressed in microns measured by a standard short-period torsion seismometer at an epicentral distance of 100km. It is one of \textit{"the most important and also the most difficult parameters"} involved in real-time seismology \cite{mag_EEW} particularly since most magnitude scales such as local magnitude ($m_L$), body wave magnitude ($m_B$), surface wave magnitude ($m_S$) are empirical and saturate at different magnitude ranges\cite{empirical1, empirical2}. This, coupled with the complexity of the nature of the geophysical processes affecting earthquakes, makes it very difficult to have a single reliable measure for the size of an earthquake\cite{Kanamori}. Magnitude values measured in different scales may thus differ by more than 1 unit, particularly for extremely large events due to saturation effects \cite{sat1, sat2, sat3, sat4}. Even for the same magnitude scale, values reported by different agencies may differ by up to 0.5 units\cite{Mousavi}. Traditionally, frequency-domain parameters such as predominant period $\tau_ p^{max}$\cite{Nakamura,allen}, effective average period $\tau_c$ \cite{kanamori2005, mag_estimation, mag_EEW} and amplitude domain parameters such as peak displacement ($P_d$) \cite{wuzhao,mag_estimation, mag_EEW} calculated from the initial 1-3 seconds of P-waves have been shown to provide reliable estimates of (body wave) magnitudes through empirical relations. Such methods have been applied to Earthquake Early Warning (EEW) systems in Japan, California, Taiwan etc. (\cite{EEW1} and the references therein). It has further been shown that the correlation of such parameters increases steadily upon increasing the duration of data used \cite{ziv}. Thus, there is an \textit{“inherent trade-off between speed and reliability”} \cite{Meier}.

Traditional machine-learning algorithms were "\textit{limited by their inability to process data in its raw format}"\cite{DL} and the need for hand-crafted features. This challenge has been overcome by the emergence of deep learning. Deep learning comprises hierarchical feature learning methods \cite{DL}, whereby several simple non-linear mathematical functions are applied to the raw data, to extract an increasingly abstract representation of the data at each level. It is the job of the deep learning model to learn the parameters of these functions. The advent of deep learning, coupled with the availability of large volumes of data and affordable computational power in the form of GPUs, have led to state-of-the-art results in image recognition \cite{imagenet,imrec,imrec2,imrec3}, speech recognition \cite{speech1,speech2,speech3}, and natural language processing \cite{nlp1,nlp2,nlp3,nlp4}. In fields such as seismology, which have been data-intensive since their very origin and are witnessing an exponential increase in the volume of data \cite{kong}, deep learning has proven successful in several tasks such as event detection\cite{preseis, wang, det1, det2, det3, det4, perol, li, Meier} and phase picking \cite{pick, det4, liao2021arru, epick}, event location characterisation \cite{perol,loc2,loc3}, first motion polarity detection \cite{polarity}, among others.

A deep learning based approach for magnitude estimation was presented by  \cite{Mousavi}. The model presented in that paper focuses on estimating the magnitude for an earthquake waveform, using a window length of 30 seconds that includes both the P- and S-wave information. The input to the model are earthquake traces, and event-vs-noise discrimination or first P-arrival are not included in its goals. The use of deep learning facilitates the learning of the most relevant features directly from the waveform. This approach suffers from the drawback of under-estimation at high magnitudes as these magnitudes are rare in nature and, hence, under-represented in the training data. In order to overcome this drawback we propose a two-pronged approach -- resampling the data to get a more uniform magnitude distribution and penalising the underestimation of high magnitudes during model training. The model presented in \cite{Mousavi} is in itself not suitable for the purpose of rapid characterisation and EEW, which takes advantage of the fact that the P-waves travel faster than the more devastating S-waves and surface waves \cite{eew3}, as that it needs the S-wave information in order to make a estimation.

In this paper we present a novel approach to achieve multi-tasking Convolutional Recurrent model for Earthquake Identification and Magnitude Estimation (CREIME), which can simultaneously perform earthquake identification, local magnitude estimation and first P-arrival time regression solely based on 1-2 seconds P-wave recording. Unlike \cite{ML_mag} which uses a set of twelve features extracted from 3 seconds of data to perform magnitude estimation, CREIME is end-to-end using a combination of Convolutional and Recurrent neural network to extract features directly from the raw waveform. The motivation for using such a small duration of P-wave data lies in its potentially easier utility in applications such as rapid earthquake characterisation for EEW systems (\cite{EEW1}\cite{EEW2} and references therein). While multiple-station based approaches are generally more robust and reliable, single station approaches are faster and therefore can be more useful in places where human settlements may lie very close to the earthquake epicenter, such as Southern California.

The local magnitude or Richter scale magnitude \cite{richter} has the form: 
\begin{equation}
M_L = \log A-\log A_0 + S 
\end{equation}
where, $A$ is the peak horizontal amplitude measured on a Wood-Anderson seismograph, and $A_0$ and $S$ are empirically determined distance and station correction terms derived from amplitude-distance relations representing attenuation and site functions respectively. While the peak amplitude can be directly obtained from the input data as we do not apply normalisation, it is expected that the model will learn the distance parameters, which are not provided explicitly, from the frequency content of the data itself. All three components are provided, to facilitate the learning of site effects (a similar approach has been followed by \cite{Mousavi}). We provide the data in units of ‘counts’ and do not perform instrument corrections, which gives the advantage that the analysis can be done in real-time. 

We demonstrate the robustness of our model, by testing it on two datasets. It is ensured that these datasets have no overlap in terms of the traces they contain to assert the generalizability of the model. We also compare the effects of using different types of ground motion data as the input to the model. As a final step we also test the model on S-arrivals which are not encountered by the model during training, to verify that S-arrivals from low magnitude events do not get wrongly identified as P-arrivals for high magnitude events. This implies that the model can be directly used on real time data.

\section{Data used}

\subsection{STEAD}

The data used to train and test CREIME has been obtained from the STanford EArthquake Dataset (STEAD) \cite{stead}. It is a high-quality benchmarked global dataset of labelled seismograms which have been detrended, bandpass filtered between 1.0-40.0Hz and resampled to 100Hz. There are a total of 7 different types of instruments in which the data has been recorded, of these, 99.5\% are either high-gain broad band or extremely short period. Each seismogram is of duration 1 minute. The original dataset comprises over one million 3 component (East-West, North-South and vertical) seismic traces associated with about 450,000 earthquake events and about 100,000 noise traces. The information for each trace is provided in the form of NumPy arrays \cite{numpy} of dimensions 6000 × 3. All earthquake waveforms are associated with local earthquakes with epicentral distance no greater than 350km. The metadata includes 35 attributes for each earthquake waveform and 8 attributes for each noise waveform. 

For the sake of uniformity in magnitude, of the 23 different magnitude scales in which earthquakes are reported, we only choose events for which the magnitudes are reported in the ‘ml’ scale, i.e., local magnitude as these events constitute the majority (above 70\%) of the dataset. To ensure that extremely noisy data is left out from the training and testing process only waveforms with a signal-to-noise ratio (provided in the metadata) above 10dB are used. The noise and earthquake traces are roughly divided in the ratio 60:10:30 for training, validation and test sets. For earthquake waveforms, it is made sure that all traces associated with one earthquake event are present in only one of the aforementioned three sets with the help of the ‘source\_id’ attribute from the metadata. For noise waveforms, traces corresponding to a particular station can be present in only one of the three sets. This ensures that the test dataset is “truly unseen” to the model and hence, can give a reliable evaluation of the model’s performance.

\begin{figure}
    \centering
    \includegraphics[scale = 0.6]{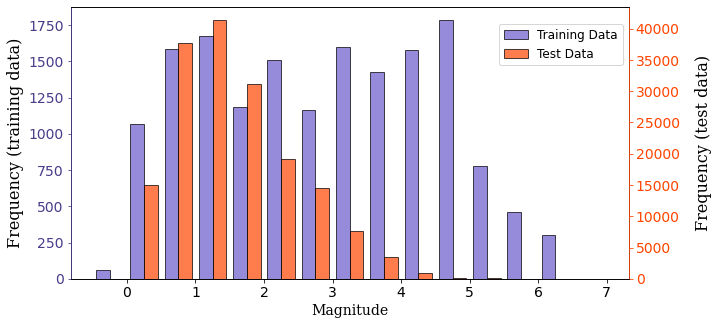}
    \caption{Distribution of magnitudes in training data (in slate blue) and chunk of STEAD\cite{stead} data used for testing (in orange). Note that the y-axis on the left corresponds to the training data distribution and that on the right corresponds to the test data distribution. While random undersampling and oversampling are applied to different magnitude ranges for training data in an attempt to get a uniform distribution, the original magnitude distribution of the test dataset is retained.} 
    \label{mag_dist}
\end{figure}

In accordance with the Gutenberg-Richter power-law \cite{GR}, high magnitude earthquakes are (fortunately) very rare in nature. This power-law is reflected in the dataset as well (with a magnitude of completeness around 1-1.5). The distribution of magnitudes in the original dataset is similar to that of the testing data  shown in Figure \ref{mag_dist}. This kind of imbalance in the distribution of the target variable in a regression problem tends to bias the model’s performance towards lower magnitudes ($<$2.5) \cite{imbalance} as observed in \cite{Mousavi}. So, to make sure that the model can perform a reliable estimation over all magnitude ranges, we perform random under-sampling up to magnitudes of 4.0 and random over-sampling for magnitudes above 4.5. For this, different rates (chosen by trial and error) of undersampling or oversampling are applied to different magnitude ranges on the training and validation sets. This results in a training set with a magnitude distribution as shown in Figure \ref{mag_dist}. No such augmentation is applied to the test set (Figure \ref{mag_dist}). Furthermore, for training and validation, the number of noise traces chosen is exactly equal to the number of event traces.

\subsection{INSTANCE}

We further test our model on the INSTANCE dataset \cite{inst}, which is a recently published dataset comprising 1.2 million three-component waveform traces and 130,000 noise traces, each with a duration of 2 minutes, recorded primarily by the Italian National Seismic Network (network code IV). Corresponding to each trace 100 metadata, including magnitude and P-wave arrival sample, are provided. To make sure that there is no overlap with the training data, we exclude data from stations that are part of the STEAD dataset. We choose only traces for which magnitudes are provided in the `ML' scale. For a fair evaluation of our model, we use only those traces with a single event and with distance and depth each within the corresponding maximum value present in the training data. Once again traces with signal-to-noise ratio lower than 10dB are not used. This leaves us with 135,347 traces corresponding to events between April, 2005 to January, 2020 and having a magnitude distribution as shown in Figure \ref{mag_inst}. The preprocessing steps for this data are very similar to those of the STEAD data except the bandpass filtering, so we apply a bandpass filter between 1.0 to 40.0 Hz using the \textit{bandpass} function from \textit{obspy.signal.filter} \cite{obspy}.

\section{Methodology}
\begin{figure}
    \begin{subfigure}{1.0\textwidth}
      \centering
      \includegraphics[width=\linewidth]{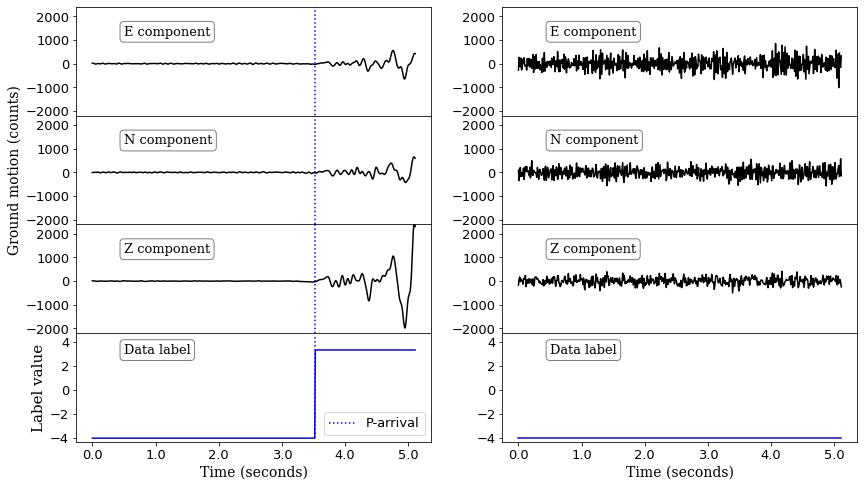}   \caption{Example of labelling for an event trace (left) and a noise trace (right); the label value is set to -4 for all samples before the P-arrival and the event magnitude for the P-arrival sample onward; for the noise trace it is set at -4 for all samples.}
      \label{lab_ev}
    \end{subfigure}
    
    \begin{subfigure}{1.0\textwidth}
      \centering
      \includegraphics[scale= 0.8, trim={0 2mm 0 0},clip]{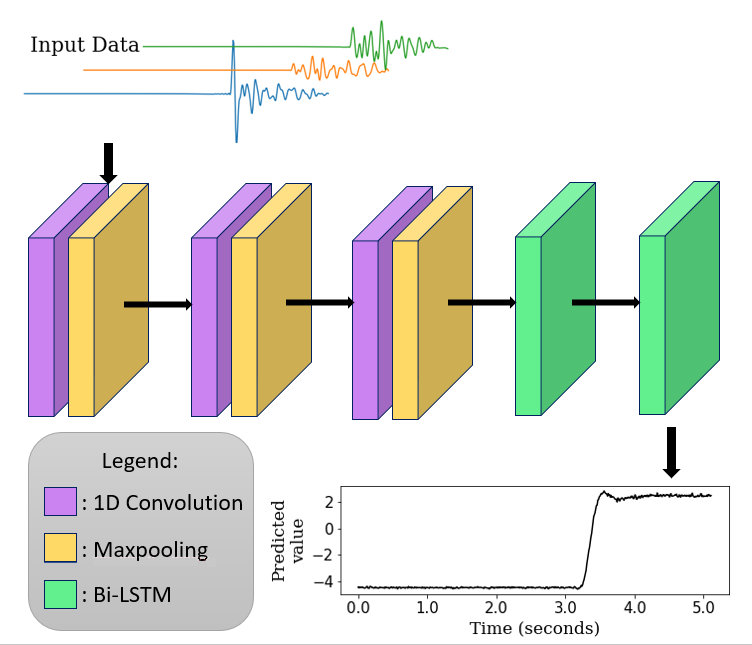}  
        
    \caption{A schematic showing the architecture of the CREIME model; each convolution layer has a kernel size 16 and the number of filters are 32, 16 and 8; each Maxpooling layer reduces the dimension of the data by a factor of 4 and the Bi-LSTM layers have dimensions of 128 and 256 respectively.}
    \label{arch}
    \end{subfigure}
    
    \caption{Data labeling and model architecture.}
\end{figure}

We use supervised learning \cite[Chapter~4]{chollet} in this work to achieve earthquake identification and magnitude estimation, together with P-arrival time regression, based upon short records of P-wave data. A sequence-to-sequence approach is developed -- the input to our model being 512 samples (5.12s) from 3 channels and the output is an array of the same length (512 samples). The data window is chosen in such a way as to include 1 to 2 seconds of P-wave data, preceded by pre-signal noise. This type of windowing allows the model to learn the noise characteristics \cite{team}. The Y-label for each X is a 512 × 1 array. These values are defined as follows:
\begin{equation}
    y_i= 
\begin{cases}
    M,& \text{if } i\geq i_p\\
    -4,              & \text{otherwise}
\end{cases}
\end{equation}
    
where $M$ is the magnitude of the event and $i_p$ denotes the P-arrival sample. The value -4 representing noise is arbitrary and chosen empirically by testing model performance on the validation data. The use of an arbitrary negative number to represent noise was explored by \cite{dpick}. An example of this labelling for event and noise data is shown in Figure \ref{lab_ev}. We have also tried modifying the final layer of the model to output two numbers corresponding P-arrival sample and magnitude instead of a sequence, similar to the approach of \cite{dpick} (not shown in the paper). However, our observation was that the sequence-to-sequence mapping approach leads to smaller errors. 

Using the conditions mentioned in the previous section we have a total of 16,178 waveform traces and an equal number of noise traces to train the model. The architecture of the CREIME model consists of three sets of 1D Convolution \cite{conv1d} and Maxpooling \cite{maxpool} layer followed by two bidirectional Long-Short Term Memory (LSTM) \cite{lstm} layers of dimensions 128 and 256, which is followed by the output layer of dimensionality 512 (Figure \ref{arch}). The convolutional and maxpooling layers are used to extract and retain the relevant features while downsampling the data volume. Bidirectional LSTMs are used because of their ability to detect temporal dependencies for sequential data such as earthquake waveforms. Each convolution has a kernel size 16, a stride of 1, and padding type “same”; the number of filters is 32, 16  and 8, respectively. Each maxpooling layer reduces the size of the data by a factor of 4. Unlike the approach in \cite{cnq} we find the model performance to be better when we use the original data without any normalisation. The model has a total of 1,454,992 trainable parameters and is trained using RMS Propagation optimiser \cite{rmsprop}, with a batch size of 512. The model is implemented using Keras \cite{keras}. On an NVIDIA A100GPU the training process takes less than 1 second per epoch. Each hyperparameter, including the number of layers in the model was chosen through meticulous experimentation by running several iterations of training and subsequent testing on the validation data.

\begin{figure}
    \centering
    \includegraphics{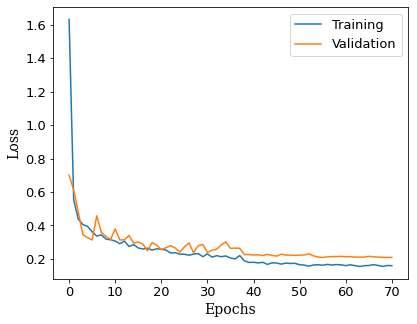}
    \caption{The variation of the training and validation loss as training progresses. The validation and training losses remain close to each other, which shows that the training is quite robust and there is no discernible overfitting.}
    \label{tr_hist}
\end{figure}

We use early stopping \cite{earlystopping} during the training to prevent overfitting. The validation loss is monitored and the training stops if it does not reduce for 15 consecutive epochs. We have an initial learning rate of $10^{-3}$ and reduce it by a factor of 10 until it reaches $10^{-6}$ if the validation loss does not go down for 10 consecutive epochs. The model with the lowest validation loss is saved. With these conditions the model trains for 71 epochs. The training history (i.e. learning curve) is shown in Figure \ref{tr_hist}.

For the cost function, we customized a combination of three losses, as different loss functions proved to be working better for different tasks and for different ranges of magnitude. The weights were determined by a trial and error method.
\begin{enumerate}
    \item Mean Squared Error (MSE) with a weight of 40\%: This is the average of squared values of errors corresponding to each data point in a minibatch. For \textit{k} output values and a batch size \textit{n} it has the form:
\begin{equation}
    \mathcal{L}_{MSE} = \frac{1}{n}\sum_{j=0}^{n-1}\frac{1}{k}\sum_{i=0}^{k-1}(y_{true}^{i,j} - y_{pred}^{i,j})^2
\end{equation}
Here $y_{true}^{i,j}$ and $y_{pred}^{i,j}$ represents the true and predicted y values of the i-th sample for the j-th example in the minibatch, respectively.
\item Mean Absolute Error (MAE) with a weight of 40\%: This is the average of absolute errors corresponding to each data point in a minibatch. For \textit{k} output values and a batch size \textit{n} it has the form:
\begin{equation}
    \mathcal{L}_{MAE} = \frac{1}{n}\sum_{j=0}^{n-1}\frac{1}{k}\sum_{i=0}^{k-1}|y_{true}^{i,j} - y_{pred}^{i,j}|
\end{equation}

\item Magnitude Estimation Loss with a weight of weight: 20\%: As mentioned in section 1, we penalise the underestimation of magnitude, for high magnitude events (and overestimation for noise traces). To achieve this we define a third loss function. For \textit{k} output values and a batch size \textit{n} it has the form:
\begin{equation}
    \mathcal{L}_{ME} = \frac{1}{n}\sum_{j=0}^{n-1}\alpha^j \frac{1}{k}\sum_{i=0}^{k-1}(y_{true}^{i,j} - y_{pred}^{i,j})
\end{equation}
where, \[
    \alpha^j= 
\begin{cases}
    \text{Event Magnitude},& \text{for events}\\
    -4,              & \text{for noise}
\end{cases}
\]
\end{enumerate}

\section{Results}
We utilise the output from our model to perform three tasks: 
discrimination between seismic event and noise, magnitude estimation,
P-arrival sample detection.

Based on a manual investigation of the output data and a subsequent testing on the validation dataset we used the following analysis to extract the desired parameters from the 512 sample sequence output by the model:

\begin{enumerate}[label=(\roman*)]
    \item Predicted magnitude, \begin{equation}\label{mpred}
        M_{pred} = \frac{1}{10}\sum_{i = (k - 9)}^k y^i_{pred}
    \end{equation} where k is the number of samples, in our case, 512
    \item Considering the first sample point in the data window as zeroth sample, we define \begin{equation} \text{P-arrival sample} =
    i^p_{pred} \text{ such that } y^i_{pred} > -0.5 \text{ for all } i \geq i^p_{pred} 
    \end{equation} \item If $M_{pred}$ calculated by equation (\ref{mpred}) is less than -0.5 then it is classified as noise. 
\end{enumerate}
 
The value -0.5 is chosen empirically based on the magnitude range of the data. For a detailed description of the metrics please refer to the Appendix A.

\subsection{Comparison with other models}

We compare our model with ones published in the papers listed below. It is important to note here, that the input data for the models in these studies differs from our data in terms of length, pre-processing etc. Therefore, for an unbiased comparison, all models are trained on the same data. In essence this is a comparison between different architectures and not between the final trained models presented by the respective authors.
\begin{enumerate}
    \item MagNet \cite{Mousavi}: This paper presents a deep learning model to perform only magnitude estimation using 30 seconds of data including both P and S phases. While both MagNet and CREIME use a combination of CNNs and bidirectional LSTMs, they differ significantly in the number of layers (MagNet uses 2 Convolutional layers and 1 bi-LSTM whereas CREIME uses 3 Convolutional layers and 2 bi-LSTMs), the model output(MagNet outputs the estimated magnitude and the aleatoric uncertainty whereas CREIME outputs a 512 dimensional array) and the choice of hyperparameters (such as number of filters and dimension of LSTM). Unlike MagNet, CREIME does not use dropout layers. The only modification we make to the original architecture of MagNet, while re-training it, is to change the input shape from (3000,3) to (512,3). We then compare this model with CREIME in terms of estimation of event magnitudes.
    
    \item CNN model for signal noise discrimination \cite{Meier}: The model presented in this paper originally takes 4s of data, starting 0.5 to 1.5 seconds before the P-arrival to discriminate between earthquake signals and noise. We train it on our data while keeping the architecture intact except a change in the input dimensions. Unlike the original paper, however, we do not impose a lower limit on the magnitudes of the events.
    
    \item ConvNetQuake\_INGV \cite{cnq}: This model is inspired by the ConvNetQuake \cite{perol}, and uses 10 seconds of data to perform multiclass classification to identify seismic events and characterise earthquake parameters such as magnitude, distance, depth and azimuth. While the original architecture uses 9 CNN layers, each downsampling the data by a factor of 2, we use only 8 (similar to \cite{perol}) since the length of data in our case is almost half of that in the original paper. Further, in the last layer we use 31 classes for magnitude instead of 20 in the original paper giving a total of 32 nodes (one for signal vs noise discrimination). To compare the magnitude regression performance with CREIME we take the predicted magnitude to be the arithmetic mean of the boundaries for the predicted class.
\end{enumerate}

In addition to these deep learning model, we also compare our model with the Short-Term Average/Long-Term Average method (STA/LTA)\cite{stalta}, to evaluate the performance of our model in terms of classification and P-arrival time prediction. This is done by using the \textit{classic\_sta\_lta} from \textit{Obspy} \cite{obspy}. The best set of parameters, determined on the basis of a grid search on the training data are: short-term window length = 20 samples (0.2s),  long-term window length = 200 samples (2s), and trigger threshold = 4.0.

\subsection{Model Performance}

\begin{figure}
    \begin{subfigure}{.5\textwidth}
      \centering
      \includegraphics[width=.9\linewidth]{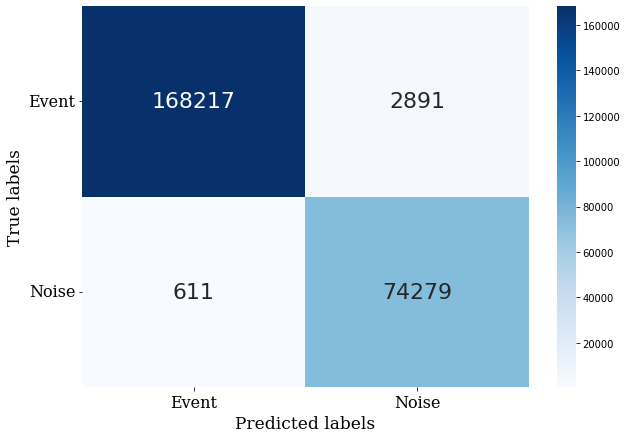}  
      \caption{Confusion matrix for classifier performance}
      \label{cm}
    \end{subfigure}
    \begin{subfigure}{.5\textwidth}
      \centering
      \includegraphics[width=.9\linewidth]{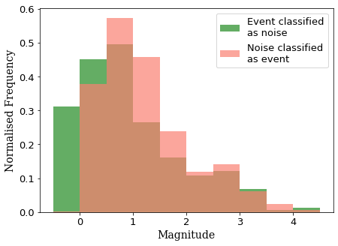}  
      \caption{Distribution of predicted magnitudes of noise misclassified as event and true magnitude of events misclassified as noise}
      \label{mag_misc}
    \end{subfigure}
    
    \caption{Analysis of CREIME model performance as a classifier on STEAD Data. It achieves an accuracy of 98.58\%. The true magnitude of events misclassified as noise and predicted magnitude of events misclassified as noise tends to be low, which reduces the chance of missed or false alarms.}
    \label{fig:fig}
\end{figure}

\begin{figure}
    \begin{subfigure}{1.0\textwidth}
      \centering
      \includegraphics[width=\linewidth]{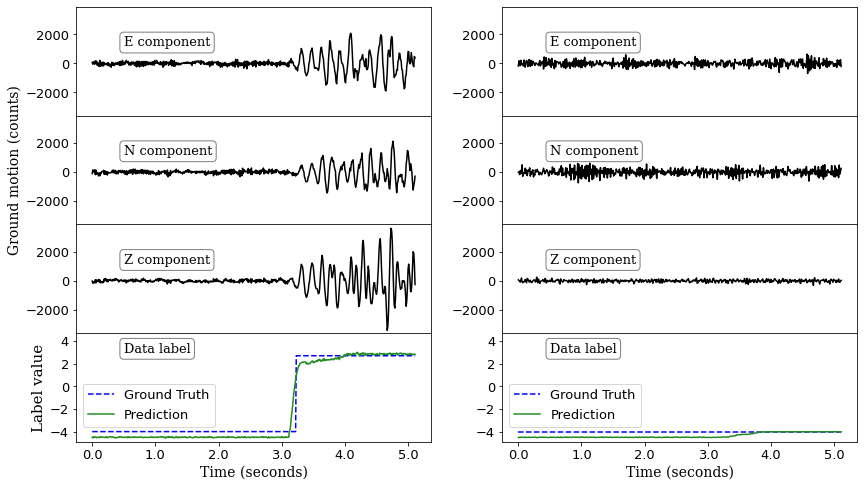}   
      \caption{Example of correct classification of an event trace (left) and a noise trace (right); one can see that the predicted magnitude for the event trace is very close to the true magnitude.}
      \label{corr}
    \end{subfigure}

    \begin{subfigure}{1.0\textwidth}
      \centering
      \includegraphics[width=\linewidth]{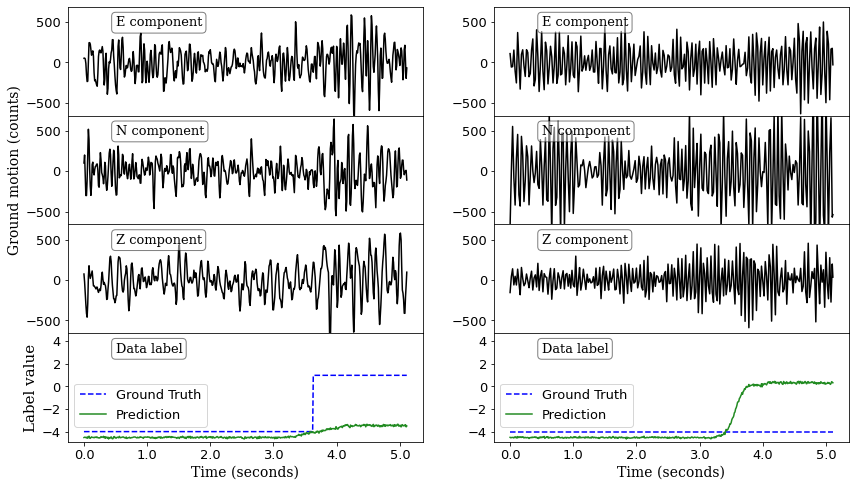}   
      \caption{Example of incorrect classification of an event trace as noise trace(left) and a noise trace as an event trace (right); the event is a low magnitude one, and quite difficult to identify in this frequency range; the noise level in case of the noise trace is quite high gets classified as a low magnitude earthquake.}
      \label{corr}
    \end{subfigure}
    \caption{Examples of true and predicted labels for test data}
    \label{examples}

\end{figure}

The model was tested on a chunk of the STEAD dataset (Figure \ref{mag_dist}). Figure \ref{cm} shows the confusion matrix for noise/event classification; Figure \ref{mag_misc} shows that the predicted magnitudes for noise data wrongly classified as event tend to be low (mostly $\leq$ 1) indicating that the possibility of false alarms caused by noise is low and events which are wrongly classified as noise are usually of low magnitude ($\leq$ 2) indicating a low risk of false alarms, which is reassuring. Figure \ref{examples} shows examples of input and corresponding outputs for correctly and incorrectly classified traces.

\begin{figure}[t]
    \begin{subfigure}{\textwidth}
      \centering
      \includegraphics[width=.7\linewidth]{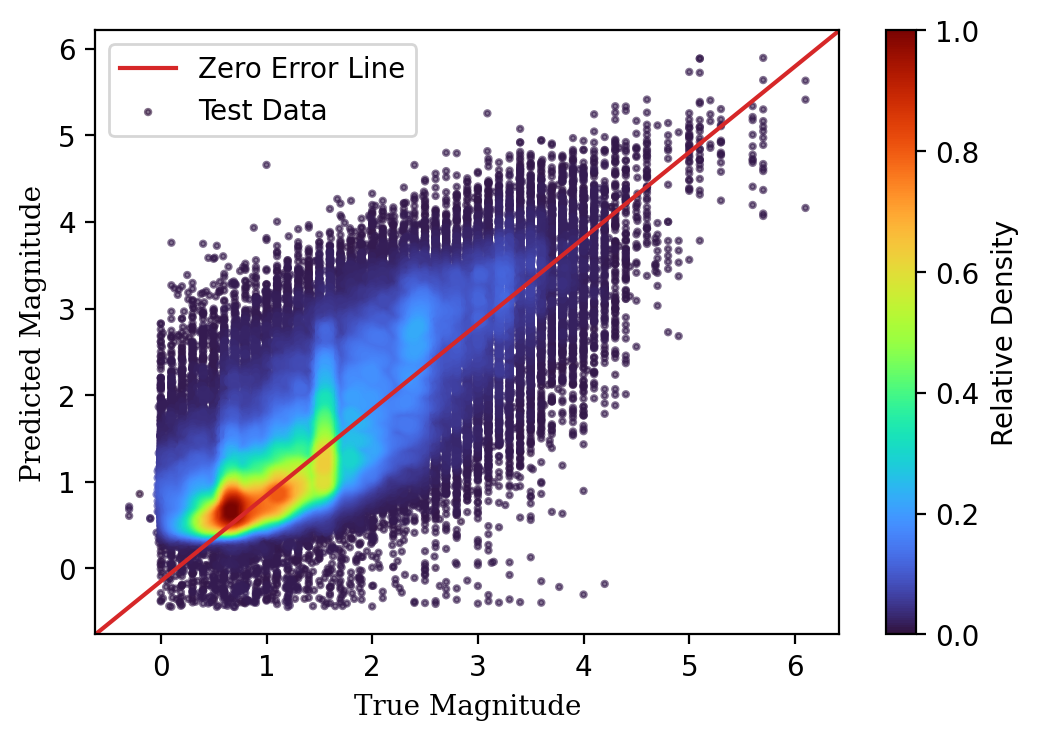}  
      \caption{Relationship between true and predicted magnitude values}
      \label{mag_scatter}
    \end{subfigure}
    \newline
    
    \begin{subfigure}{.5\textwidth}
      \centering
      \includegraphics[width=\linewidth]{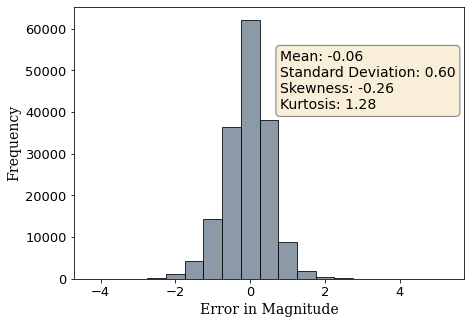}  
      \caption{Distribution of errors in predicted magnitudes}
      \label{mag_error}
    \end{subfigure}
    \begin{subfigure}{.5\textwidth}
      \centering
      \includegraphics[width=\linewidth]{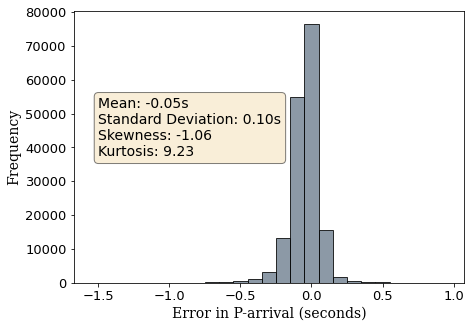}  
      \caption{Distribution of errors in P-arrival estimation}
      \label{P_error}
    \end{subfigure}
    \newline
    \caption{Analysis of model performance as a regressor on STEAD Data. The density plot shows that the highest density of points lies close to the zero error line; in spite of our penalization of under-estimation of high magnitudes, some under-estimation is observed above a magnitude of 5.5. In over 90\% of the cases, the error in predicted magnitudes is less than 1 unit.}
    \label{fig:reg_stead}
\end{figure}

The scatter plot for predicted versus true magnitudes is shown in Figure \ref{mag_scatter}. It is worth to note that for majority of the events (shown with higher relative density in the plot) the prediction reproduce well for the true magnitudes up to 5.5. For higher magnitudes events, some degree of underestimation is observed indicating one may further strengthen the penalty for such prediction into the loss designing. 
The result here is still an improvement over \cite{Mousavi}, where magnitude underestimation starts to occur from a magnitude of 4. The histogram for errors in magnitude (Figure \ref{mag_error}) has a mean of -0.06 units, and a slight left skew, reflecting our penalisation of underestimation of magnitudes. The histogram for errors in predicted P-arrival (Figure \ref{P_error}), is also unimodal, with a higher negative skewness indicating, that the P-arrival is more often predicted to be at a \textit{later} time than it really is. The kurtosis for errors in P-arrival prediction is also much higher than that for magnitude prediction, indicating that errors in P-arrival predictions have a much narrower peak compared to errors in magnitude prediction. Similar results are observed for the INSTANCE dataset. We refer interested readers to Appendix B for the corresponding figures.

The performance metrics for the CREIME classifier in comparison with other classification models and the STA/LTA algorithm are summed up in Table \ref{tab:class}. CREIME outperforms all the other architectures trained on the same data, and the conventional STA/LTA algorithm. The performance of the model in estimating magnitude and P-arrival time is summarised in tables \ref{tab:reg1} and \ref{tab:reg2} respectively. CREIME model outperforms MagNet and ConvNetQuake\_INGV in terms of magnitude estimation. It also gives lower values for both RMSE and MAE compared to STA/LTA algorithm. 
 
\begin{table}
    \centering
    \caption{Comparison between the performance of CREIME model and other baseline models as a classifier for events and noise. CREIME model outperforms the other models.}
    \scriptsize
    \begin{tabular}{llccccccc}
 
& & \multicolumn{7}{c}{\textbf{\textit{Metric}}} \\
 \cline{3-9}
 & &\multirow{2}{*}{\textbf{Accuracy (\%)}} & \multicolumn{2}{c}{\textbf{Precision (\%)}} & \multicolumn{2}{c}{\textbf{Recall (\%)}} & \multicolumn{2}{c}{\textbf{F1-score (\%)}}\\ 
 \cline{4-9}
\textbf{\textit{Dataset}} & \textbf{\textit{Model Architecture}} & & Event & Noise & Event & Noise & Event & Noise \\
 \hline\hline  \\[-2ex]
\multirow{4}{*}{\textbf{STEAD}} & \textit{CREIME Model} & \textbf{98.58}
& \textbf{99.64} & \textbf{96.25} & \textbf{98.31} & \textbf{99.18} & \textbf{98.97} & \textbf{97.70} \\
&\textit{CNN Model} & 89.72
& 99.18 & 75.37 & 85.93 & 98.37 & 92.08 & 85.35 \\
&\textit{ConvNetQuake\_INGV} & 96.56 & 99.12 & 91.30 & 95.91 & 98.05 & 97.49 & 94.55 \\
&\textit{STA/LTA Algorithm} & 94.08 
& 96.03 & 89.70 & 95.43 & 91.00 & 95.73 & 90.34 \\
\hline \\[-2ex]
\multirow{4}{*}{\textbf{INSTANCE}} &
\textit{CREIME Model} & \textbf{97.59}
& \textbf{98.66} & \textbf{95.75} & \textbf{97.53} & \textbf{97.68} & \textbf{98.10} & \textbf{96.71} \\
&\textit{CNN Model} & 91.71
& 96.77 & 84.33 & 90.00 & 94.71 & 93.23 & 89.22 \\
&\textit{ConvNetQuake\_INGV} & 86.48 & 96.00 & 74.16 & 82.79 & 93.47 & 88.90 & 82.70 \\
&\textit{STA/LTA Algorithm} & 86.03 
& 90.87 & 78.49 & 86.81 & 84.66 & 88.79 & 81.46 \\
\hline
    \end{tabular}
    \label{tab:class}
\end{table}

\begin{table}
    \centering
    \caption{Comparison between magnitude estimation by CREIME model and other baseline models. The smallest errors are shown by CREIME model.}
    \scriptsize
    
    \begin{tabular}{llcccc}
 
& & \multicolumn{4}{c}{\textbf{\textit{Metric}}} \\
 \cline{3-6}
& & \textbf{Mean Error} & \textbf{St. dev. of Error} & \textbf{RMSE} & \textbf{MAE}\\ 
 
\textbf{\textit{Dataset}} & \textbf{\textit{Model Architecture}}  \\
 \hline\hline  \\[-2ex]
 \multirow{3}{*}{\textbf{STEAD}} & \textit{CREIME Model} & \textbf{-0.06} & \textbf{0.60} & \textbf{0.61} & \textbf{0.46} \\
&\textit{MagNet} & -0.29 & 0.65 & 0.72 & 0.53\\
&\textit{ConvNetQuake\_INGV} & 0.41 & 1.05 & 1.13 & 0.94 \\
\hline \\[-2ex]
\multirow{3}{*}{\textbf{INSTANCE}} & \textit{CREIME Model} & \textbf{-0.02} & \textbf{0.69} & \textbf{0.69} & \textbf{0.54} \\
& \textit{MagNet} & -0.33 & 0.80 & 0.86 & 0.68\\
& \textit{ConvNetQuake\_INGV} & 0.78 & 0.98 & 1.25 & 1.04 \\
\hline
    \end{tabular}
    
    \label{tab:reg1}
\end{table}

\begin{table}
    \centering
    \caption{Comparison between CREIME model and STA/LTA method in terms of P-arrival picking. CREIME model outperforms STA/LTA}
    \scriptsize
    
    \begin{tabular}{llcccc}
 
& & \multicolumn{4}{c}{\textbf{\textit{Metric}}} \\
 \cline{3-6}
 & & \textbf{Mean Error} & \textbf{St. dev. of Error} & \textbf{RMSE} & \textbf{MAE}\\ 
 
\textbf{\textit{Dataset}} & \textbf{\textit{Model Architecture}} & \textbf{(s)} & \textbf{(s)} & \textbf{(s)} & \textbf{(s)}  \\
 \hline\hline  \\[-2ex]
 \multirow{2}{*}{\textbf{STEAD}} &\textit{CREIME Model} & -0.05 & \textbf{0.10} & \textbf{0.12} & \textbf{0.08} \\
&\textit{STA/LTA} & \textbf{0.01} & 0.37 & 0.36 & 0.18\\
\hline\\[-2ex]
\multirow{2}{*}{\textbf{INSTANCE}}&\textit{CREIME Model} & -0.04 & \textbf{0.13} & \textbf{0.14} & \textbf{0.09} \\
&\textit{STA/LTA} & \textbf{0.01} & 0.52 & 0.52 & 0.29\\
\hline
    \end{tabular}
\label{tab:reg2}
\end{table}

\section{Discussion}

\subsection{Factors affecting the model performance}
\begin{figure}
    \begin{subfigure}{.5\textwidth}
      \centering
      \includegraphics[width=\linewidth]{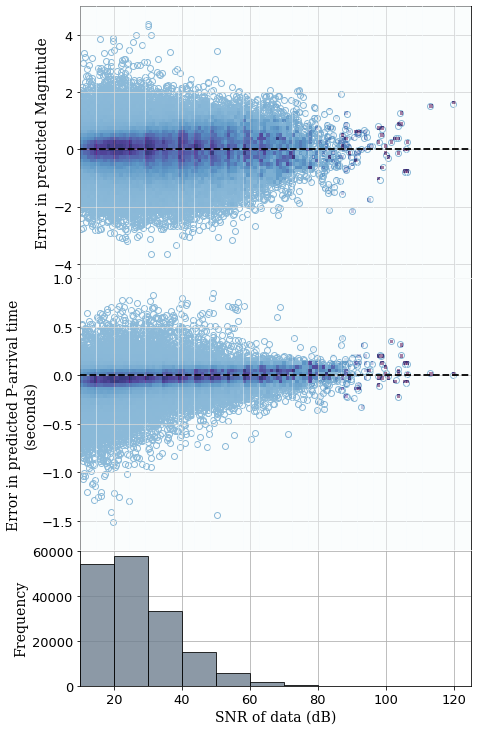}  
        \caption{Variation of errors with signal to noise ratio}
        \label{err_snr}
        
    \end{subfigure}
    \begin{subfigure}{.5\textwidth}
      \centering
      \includegraphics[width=\linewidth]{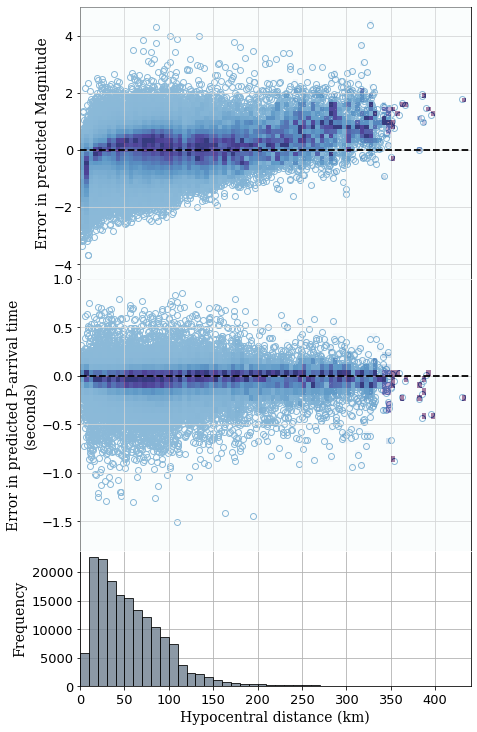}  
        \caption{Variation of errors with hypocentral distance}
         \label{err_dist}
       
    \end{subfigure}
    \caption{Factors affecting the error in estimation of magnitude and P-arrival times; errors in both magnitude and P-arrival are lower for higher signal-to-noise ratios; the magnitude of events seems to be under-estimated for higher hypocentral distances owing to their under-representation in the data.}
    \label{fig}
\end{figure}

We investigated the different factors that influence the results of our model. Figure \ref{err_snr} shows the variation of errors with the signal-to-noise ratio in the data. It is observed that the errors in magnitude and P-arrival time show highest density within $\pm 1$ units and $\pm 0.1$ seconds, respectively, and tends to be lower for higher signal-to-noise ratios. 

Figure \ref{err_dist} shows the variation of errors with hypocentral distance. We see that the errors tend to be close to zero over a wide range of hypocentral distances (up to 200km). There is a tendency for the model to underestimate the magnitude for higher hypocentral distances, which are under-represented in the training data. 

Both these figures are generated using STEAD data, and the corresponding figures for INSTANCE data can be found in the Appendix B.

\subsection{Effect of using different types of ground motion data as input}
We also compared the performance of the model when trained on different kinds of ground motion data viz. acceleration (in $\mu$m s$^{-2}$), velocity (in nm s$^{-1}$) and displacement (in nm) to investigate the effects of instrument response removal. This part of the analysis was done only on the STEAD data. We lose roughly one fourth of the data due to unavailability of the instrument response. In each case, the models were trained on roughly the same number of traces, alongside which we also compare it with the model whose results are discussed in the previous section (trained on more traces).  For a fair comparison, we also train a model on raw data, using roughly the same number of traces as in case of ground motion data, accounting for the loss of data due to unavailability of instrument response (this model is referred to as raw data (smaller) in the tables). The reason behind doing this is to highlight one of the advantages of using counts data without instrument response removal, which is the availability of more traces for training and testing. All five models have been tested on the same traces. 

Tables \ref{tab:class_gm}-\ref{tab:reg2_gm} show the comparison between different types of input data. Even though certain ground motion parameters perform better in some metrics, using the raw data gives us an advantage that the data can be used in real time, and it is much readily available.

To make sure, that S-arrivals for low magnitude earthquakes do not get detected as high-magnitude events, we test the model on S-arrival data. We do not notice any systematic overestimation, only in 9\% of the cases in the overestimation more than 1 unit. This means that our model can be applied to the incoming seismogram in real time for rapid characterisation. Comparing the performance of the CREIME model with our observations in \cite{classifier}, we find that providing data labels in the form of a series and including the first P-arrival information is beneficial for the model, in estimating the earthquake magnitude.

\begin{table}
    \centering
    \caption{Summary of classification performance for different types of ground motion data; removing instrument response does not seem to provide a significant advantage over using raw data}
    \scriptsize
    
    \begin{tabular}{lccccccc}
 
 & \multicolumn{7}{c}{\textbf{\textit{Metric}}} \\
 \cline{2-8}
 &\multirow{2}{*}{\textbf{Accuracy (\%)}} & \multicolumn{2}{c}{\textbf{Precision (\%)}} & \multicolumn{2}{c}{\textbf{Recall (\%)}} & \multicolumn{2}{c}{\textbf{F1-score (\%)}}\\ 
 \cline{3-8}
\textbf{\textit{Type of input data}} & & Event & Noise & Event & Noise & Event & Noise \\
 \hline\hline  \\[-2ex]
 \textit{Raw Data} & \textbf{98.33}
& 99.90 & 85.90 & \textbf{98.25} & 99.09 & \textbf{99.06} & 92.03 \\
 \textit{Raw Data (smaller)} & 97.85
& \textbf{99.91} & 82.36 & 97.71 & 99.17 & 98.79 & 89.99 \\
\textit{Acceleration} & 98.15
& 99.79 & \textbf{93.62} & 97.75 & \textbf{99.37} & 98.76 & \textbf{96.41} \\
\textit{Velocity} & 97.81 & 99.65 & 92.74 & 97.42 & 98.98 & 98.53 & 95.76 \\
\textit{Displacement} & 96.52 
& 99.61 & 88.54 & 95.74 & 98.86 & 97.64 & 93.41 \\
\hline
    \end{tabular}
    \label{tab:class_gm}
\end{table}

\begin{table}
    \centering
    \caption{Summary of magnitude estimation for different types of ground motion data}
    \scriptsize
    
    \begin{tabular}{lcccc}
 
 & \multicolumn{4}{c}{\textbf{\textit{Metric}}} \\
 \cline{2-5}
 & \textbf{Mean Error} & \textbf{St. dev. of Error} & \textbf{RMSE} & \textbf{MAE}\\ 
 
\textbf{\textit{Ground Motion}}  \\
 \hline\hline  \\[-2ex]
 \textit{Raw Data} & -0.19 & 0.63 & 0.65 & 0.50 \\
 \textit{Raw Data (smaller)} & \textbf{0.01} & 0.64 & 0.64 & 0.49 \\
\textit{Acceleration} & -0.11 & \textbf{0.56} & \textbf{0.57} & \textbf{0.44}\\
\textit{Velocity} & -0.09 & 0.62 & 0.63 & 0.47 \\
\textit{Displacement} & -0.32 & 0.65 & 0.72 & 0.54 \\
\hline
    \end{tabular}
    \label{tab:reg1_gm}
\end{table}

\begin{table}
    \centering
    \caption{Summary of P-arrival estimation for different types of ground motion data}
    \scriptsize
    
    \begin{tabular}{lcccc}
 
 & \multicolumn{4}{c}{\textbf{\textit{Metric}}} \\
 \cline{2-5}
 & \textbf{Mean Error (s)} & \textbf{St. dev. of Error (s)} & \textbf{RMSE (s)} & \textbf{MAE (s)}\\ 
 
\textbf{\textit{Ground Motion}}  \\
 \hline\hline  \\[-2ex]
 \textit{Raw Data} & \textbf{-0.04} & \textbf{0.11} & \textbf{0.12} & \textbf{0.08} \\
 \textit{Raw Data (smaller)} & -0.06 & \textbf{0.11} & 0.13 & 0.09 \\
\textit{Acceleration} & -0.07 & 0.12 & 0.14 & 0.10\\
\textit{Velocity} & -0.06 & 0.14 & 0.15 & 0.11 \\
\textit{Displacement} & -0.06 & 0.18 & 0.19 & 0.13 \\
\hline
    \end{tabular}
    \label{tab:reg2_gm}
\end{table}

\section{Conclusions}
We present a novel deep learning model, CREIME, which successfully unifies the tasks of event and noise discrimination, P-arrival time estimation and magnitude estimation using a smaller window (up to 2 seconds) of P-wave data as compared to previously published models. The model in its current form, however, is restricted by the fact that was trained specifically on data windows where the P-wave arrival is between 3.12 and 4.12 seconds of the starting sample. This restriction can be overcome in a future version of the model by modifying the training dataset to include a wider range of arrival times. Nevertheless, this model can be seen as an important first step to a fully automated earthquake characterisation approach in real time. We show that it performs better than baseline models re-trained on the same duration of data. It also outperforms traditional event discrimination algorithms such as STA/LTA. We demonstrate the robustness of our model by testing it on two independent datasets, and show that it can provide reliable estimates over a wide range of hypocentral distances and signal-to-noise ratios. The model is designed to handle seismological waveform data in its raw format, which makes it very efficient in handling big data. Such models can also find their utility in smartphone applications to issue timely warnings to the public, as smartphone sensors have been shown to be capable of detecting seismic events \cite{myshake}. 

\section*{Acknowledgment}
This research is supported by the ``KI-Nachwuchswissenschaftlerinnen" - grant SAI 01IS20059 by the Bundesministerium für Bildung und Forschung - BMBF. Calculations were performed at the Frankfurt Institute for Advanced Studies' new GPU cluster, funded by BMBF for the project Seismologie und Artifizielle Intelligenz (SAI). The authors are also thankful to Dr. KiranKumar Thingbaijam, Dr. Jan Steinheimer and Jonas Köhler for their kind suggestions. Horst Stöcker gratefully acknowledges the Judah M. Eisenberg Laureatus - Professur at Fachbereich Physik, Goethe Universität Frankfurt, funded by the Walter Greiner Gesellschaft zur Förderung der physikalischen Grundlagenforschung e.V.

\printbibliography
\newpage
\appendix
\section{Metrics used for model evaluation}
\renewcommand{\theequation}{A.\arabic{equation}}
\setcounter{equation}{0} 
\subsection{Classification Metrics}

We use different kinds of metrics to evaluate the classification and regression tasks. The performance of a classifier is often visualised with the help of a confusion matrix \cite{Ting2017}. The metrics we use to evaluate our model performance are described below. The abbreviations used are: \begin{multicols}{2}
 TP: True positives \par
 TN: True negatives\par
 FP: False positives\par
 FN: False negatives\par
 \end{multicols}
 \begin{itemize}
     \item \textbf{Accuracy}: The accuracy of a classifier is the ratio of the number of correct predictions to the total number of predictions made by the model.
     \begin{equation}
      \text{Accuracy} = \frac{\text{TP + TN}}{\text{TP + FP + TN + FN}}   
     \end{equation}
     \item \textbf{Precision}: The precision of a classifier is the ratio of the number of correct predictions for a particular class to the total number of times that class is predicted.
     \begin{equation} \text{Precision} = \frac{\text{TP}}{\text{TP + FP}}
     \end{equation}
     \item \textbf{Recall}: The recall of a classifier is the proportion of the number of instances of a class in the data set that are correctly predicted.
     \begin{equation}
        \text{Precision} = \frac{\text{TP}}{\text{TP + FN}} 
     \end{equation}
     \item \textbf{F1 Score}: By definition, there is an inherent trade-off between the precision and the recall of a classifier. Therefore, it is often worthwhile to look at the harmonic mean of the two. This metric is called the  F1-score of the classifier.
     \begin{equation}
       \text{F1-score} = \frac{2 \times \text{Precision }\times \text{ Recall}}{\text{Precision + Recall}}  
     \end{equation}
 \end{itemize} 
\subsection{Regression Metrics}
For the regression task, the following metrics will be used to measure the CREIME performance:
\begin{itemize}
    \item \textbf{Mean Error}: This is the mean value of errors corresponding to each example in the data set.
    \begin{equation}
        \text{Mean Error, } \Bar{\mathcal{E}}= \frac{1}{N}\sum_{i = 0}^{N-1} \mathcal{E}_i = \frac{1}{N}\sum_{i = 0}^{N-1} y_{true}^i - y_{pred}^i
    \end{equation}
    where N is the total number of examples in the dataset.
    \item \textbf{Standard Deviation of Error}: This is the standard deviation of the errors in the predictions. \begin{equation}
        \text{Standard Deviation of Error, } \sigma_\mathcal{E} = \sqrt{\frac{\sum_{i = 0}^{N-1}(\mathcal{E}_i - \Bar{\mathcal{E}})^2}{N}}
    \end{equation}
    \item \textbf{Root Mean Squared Error (RMSE)}: As the name says, this is the square root of the mean of squares of errors in prediction.
    \begin{equation}
        \text{RMSE} = \sqrt{\frac{\sum_{i = 0}^{N-1}\mathcal{E}_i^2}{N}}
    \end{equation}
    \item \textbf{Mean Absolute Error}: This is the mean of the absolute values of the errors in prediction.
    \begin{equation}
        \text{MAE} = \frac{\sum_{i = 0}^{N-1}|\mathcal{E}_i|}{N}
    \end{equation}

\end{itemize}

\newpage

\section{Model performance on INSTANCE Dataset}

\renewcommand{\thefigure}{B.\arabic{figure}}
\setcounter{figure}{0} 

\begin{figure}[!htb]
    \centering
    \includegraphics[scale = 0.6]{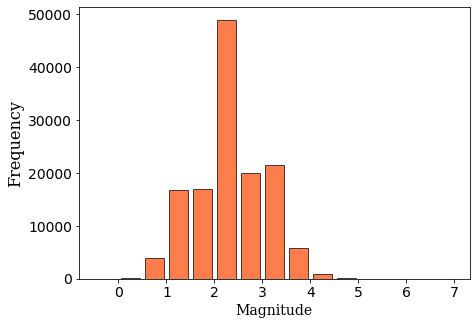}
    \caption{Distribution of magnitudes in chunk of INSTANCE data used for testing. Once again, no resampling is applied to the dataset based on magnitude.} 
    \label{mag_inst}
\end{figure}

\begin{figure}[!htb]
\begin{subfigure}{.5\textwidth}
  \centering
  \includegraphics[width=.9\linewidth]{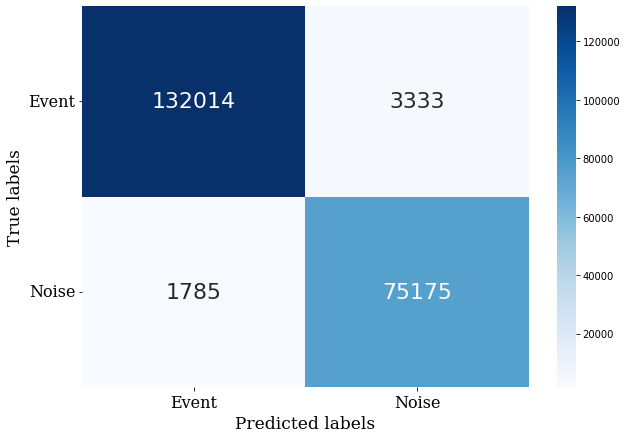}  
  \caption{Confusion matrix for classifier performance on Instance Data.}
  \label{cm_inst}
\end{subfigure}
\begin{subfigure}{.5\textwidth}
  \centering
  \includegraphics[width=.9\linewidth]{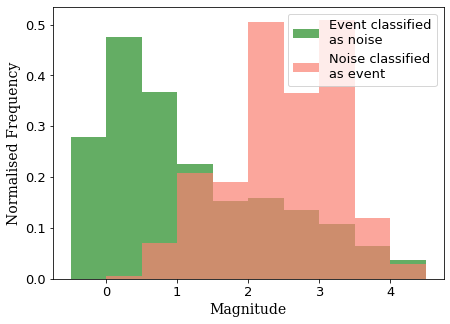}  
  \caption{Distribution of predicted magnitudes of noise misclassified as event and true magnitude of events misclassified as noise.}
  \label{mag_nase_inst}
\end{subfigure}

\caption{Analysis of model performance as a regressor on INSTANCE Data, here the events misclassified as noise, reflect the imbalanced distribution of magnitudes in the dataset itself, whereas the predicted magnitude of noise waveforms follows a similar trend as in case of STEAD data.}
\label{fig:fig}
\end{figure}

\begin{figure}[!htb]
\begin{subfigure}{\textwidth}
  \centering
  \includegraphics[width=.7\linewidth]{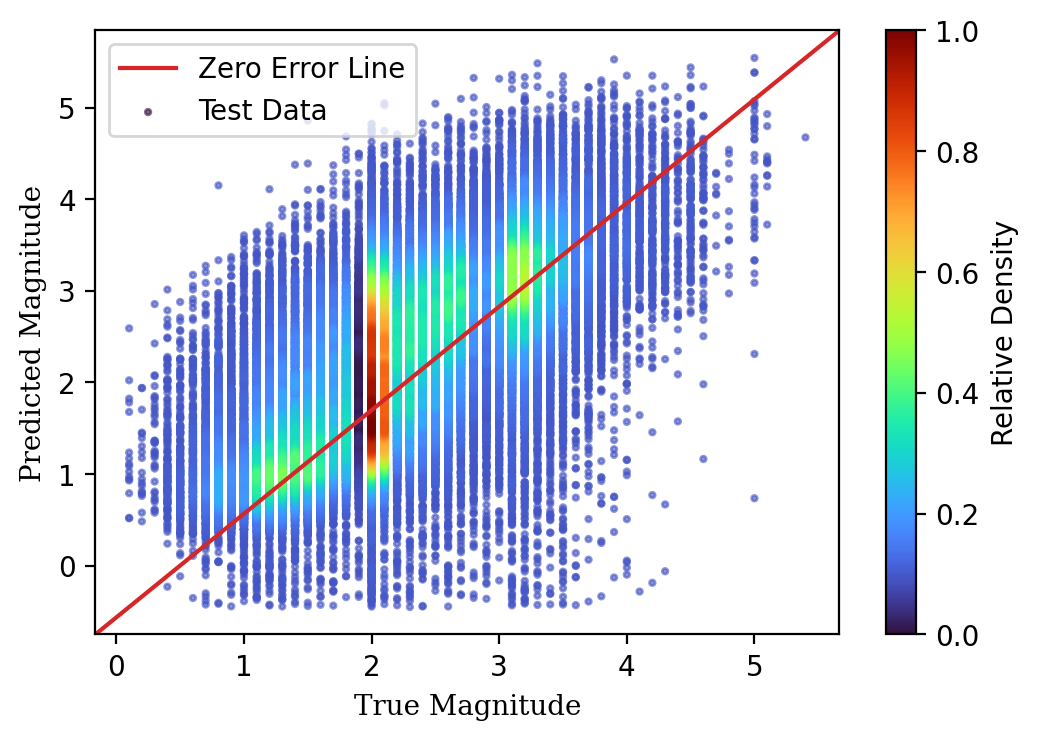}  
  \caption{Relationship between true and predicted magnitude values.}
  \label{mag_scatter_inst}
\end{subfigure}
\newline

\begin{subfigure}{.5\textwidth}
  \centering
  \includegraphics[width=\linewidth]{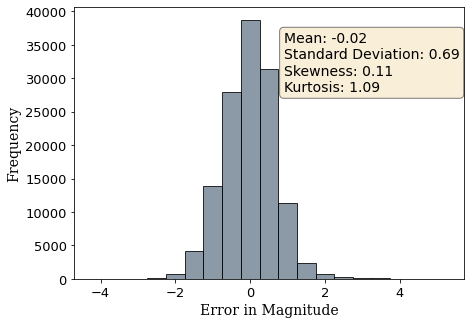}  
  \caption{Distribution of errors in predicted magnitudes.}
  \label{mag_error_inst}
\end{subfigure}
\begin{subfigure}{.5\textwidth}
  \centering
  \includegraphics[width=\linewidth]{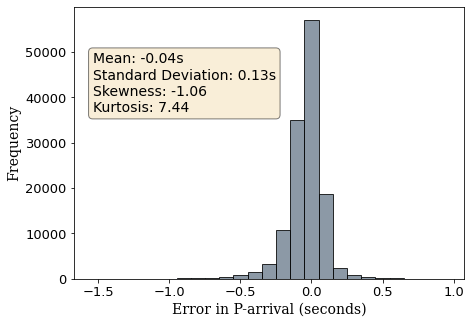}  
  \caption{Distribution of errors in P-arrival estimation.}
  \label{P_error_inst}
\end{subfigure}
\newline

\caption{Analysis of model performance as a regressor on INSTANCE Data.}
\label{fig:fig}
\end{figure}

\begin{figure}
\begin{subfigure}{.5\textwidth}
  \centering
  \includegraphics[width=\linewidth]{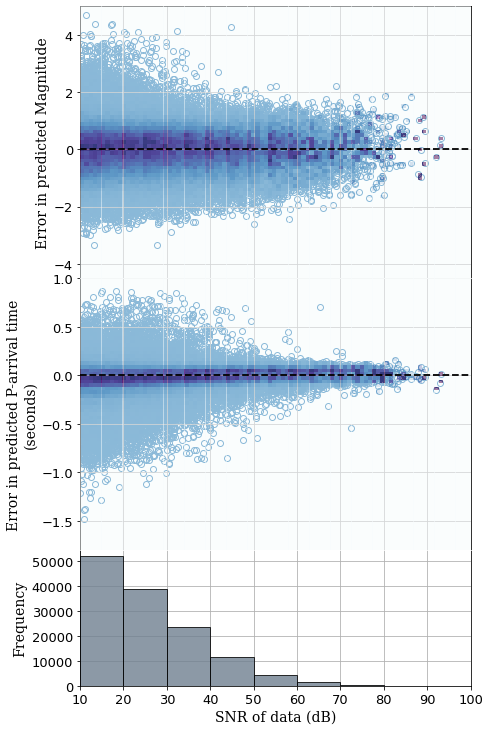}  
    \caption{Variation of errors with signal to noise ratio.}
    \label{err_snr_inst}
    
\end{subfigure}
\begin{subfigure}{.5\textwidth}
  \centering
  \includegraphics[width=\linewidth]{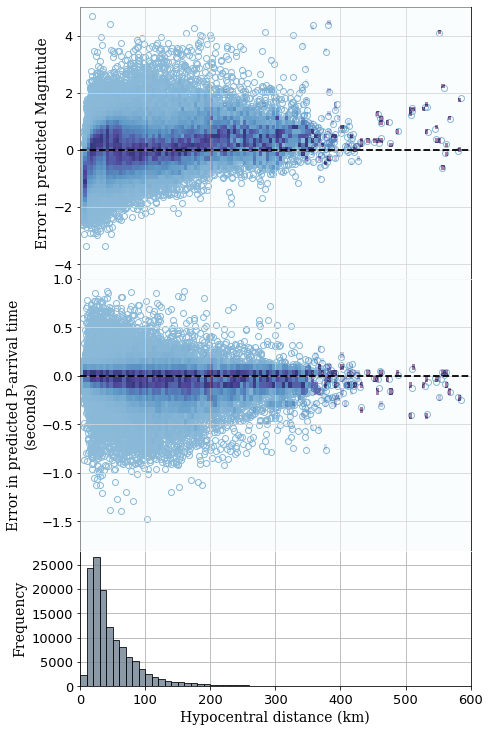}  
    \caption{Variation of errors with hypocentral distance.}
     \label{err_dist_inst}
   
\end{subfigure}
\caption{}
\label{fig}
\end{figure}

\end{document}